\documentclass[aps, pra, reprint, superscriptaddress
]{revtex4-2}
\usepackage{natbib}
\usepackage{graphicx}
\usepackage{xcolor}

\newcommand{\dt}[0]{\frac{\mathrm{d}}{\mathrm{d}t}}

\newcommand{\ii}[0]{\mathrm{i}}
\usepackage{physics}
\usepackage{amsfonts}
\usepackage{yfonts}
\begin{document}

\title{Numerically exact open quantum system work statistics with process tensors}

\author{Mike Shubrook}
\affiliation{Department of Physics and Astronomy, University of Manchester, Oxford Road, Manchester M13 9PL, United Kingdom}
\author{Moritz Cygorek}
\affiliation{Condensed Matter Theory, Technical University of Dortmund, 44227 Dortmund, Germany}
\author{Erik Gauger}
\affiliation{SUPA, Institute of Photonics and Quantum Sciences, Heriot-Watt University, Edinburgh, EH14 4AS, United Kingdom}
\author{Jake Iles-Smith}
\affiliation{School of Mathematical and Physical Sciences, University of Sheffield, Sheffield S3 7RH, United Kingdom}
\author{Ahsan Nazir}
\affiliation{Department of Physics and Astronomy, University of Manchester, Oxford Road, Manchester M13 9PL, United Kingdom}
\date{\today}

\begin{abstract}
Accurately quantifying the thermodynamic work costs of quantum operations is essential for the continued development and optimisation of emerging quantum technologies. 
This present a significant challenge in regimes of rapid control within complex, non-equilibrium environments  
--- conditions under which many contemporary quantum devices 
operate and conventional approximations break down. 
Here, we introduce a process tensor framework that enables the computation of the \emph{full numerically exact} quantum work statistics of driven open quantum systems. 
We demonstrate the utility  
of our approach by  
applying it to a Landauer erasure protocol operating beyond the weak-coupling, Markovian, and slow-driving limits. The resulting work probability distributions reveal distinct quantum signatures 
that are missed by low-order moments yet significantly impact the erasure fidelity of the protocol. 
Our framework delivers non-perturbative accuracy and detail in characterising energy-exchange fluctuations in driven open quantum systems, establishing 
a powerful and versatile tool for exploring thermodynamics and control in the operating regimes of 
both near-term and future quantum devices.
\end{abstract}

\maketitle

\section{Introduction} 
The development of practical quantum technologies  
requires precise control over quantum states to implement reliable operations while mitigating environment-induced decoherence.  
Such control inevitably entails the exchange of energy with an external 
drive in the form of thermodynamic work. 
At the quantum scale, 
characterising this work is 
more challenging than in the classical limit  
due to intrinsic quantum fluctuations and measurement back-action~\cite{talkner2007fluctuation, talkner2009fluctuation, RevModPhys.83.771,PhysRevE.94.010103,Vinjanampathy01102016,PhysRevX.6.041017,talkner2020colloquium,PhysRevLett.124.160601,PhysRevA.108.L050203}. As a result, quantum 
work is an inherently stochastic quantity 
governed by the interplay between external driving, system-environment interactions, and measurement.  
A full 
description  
therefore requires access to the 
complete work probability distribution (WPD), not just  
its low-order moments.  
Since work statistics determine fundamental bounds on work extraction, irreversibility, and entropy production,  
precise evaluation of the WPD  
is essential for advancing our understanding of quantum-scale thermodynamics (see e.g.,~\cite{PhysRevLett.78.2690,PhysRevE.60.2721,RevModPhys.81.1665,RevModPhys.83.771,Vinjanampathy01102016,Goold_2016,PhysRevA.97.062114,PhysRevE.103.052138}). However, obtaining the full WPD in realistic open quantum systems remains a major challenge, particularly under fast driving or strong environmental coupling conditions where conventional approximations fail~\cite{diba2024quantum}.

A common approach to constructing the quantum WPD is the two-point measurement protocol (TPMP), which defines work as the difference between two projective energy measurements performed before and after a driven process~\cite{talkner2007fluctuation}. While this prescription is conceptually clear, its practical application to open quantum systems is fraught with difficulties. For example, the internal energy of an open quantum system is not uniquely defined as it depends on both the system Hamiltonian and its coupling to the environment. In principle, this would seem to require consideration of an exponentially large joint Hilbert space. Perturbative approaches based on the Born-Markov or adiabatic assumptions provide approximate access to work statistics only in weak-coupling~\cite{PhysRevB.90.094304,silaev14, Liu16} or slow-driving limits~\cite{Suomela15,diba2024quantum}. Yet, many emerging quantum devices operate far beyond these regimes, where coherent control and environmental memory effects compete on similar time and energy scales~\cite{Alkauskas_2014,Wei14,Bosman17,potovcnik2018studying,Brash19,Maier2019,Clear20,Takahashi2020,Bando22,Fischer23}. 
Although recent advances in path-integral and tensor-network techniques have enabled numerically exact simulations of open system dynamics~\cite{strathearn_efficient_2018, jorgensen2019exploiting,cygorek2022ACE}, prior studies of energy exchange have typically focused either on steady-state heat transport with time-independent Hamiltonians~\cite{popovic2021quantum}, omitting the genuinely dynamical aspects of work, or specific models~\cite{campisi2009b,PhysRevB.91.224303,Aurell2017a,funo2018path}. 
As a result, 
a general and efficient means of computing the full  
non-perturbative WPD has remained elusive.

Here, we overcome these limitations by introducing a numerically exact framework for calculating the full quantum work statistics in driven open systems. Our approach builds from  
the work characteristic function (WCF), the Fourier transform of the WPD, which evolves as a pseudo density operator that propagates along both real time and an auxiliary counting field. By combining these two coordinates into a single generalised time axis, we find that the computation of work statistics becomes amenable to the process-tensor (PT) formalism~\cite{jorgensen2019exploiting,cygorek2022ACE}, which efficiently represents the Feynman–Vernon influence functional of the environment~\cite{feynman1963theory} via tensor-network compression.

The resulting framework enables the numerically exact evaluation of open quantum system WPDs in regimes of strong coupling, non-Markovianity, and non-adiabatic driving. To demonstrate this versatility, we  
consider an example of a Landauer erasure protocol~\cite{PhysRevLett.120.210601,gaudenzi_quantum_2018}  
driven at arbitrary speed in the presence of strong coupling to a bosonic environment. The computed work distributions reveal rich structure,  
including signatures of non-adiabatic coherences that adversely affect erasure fidelity and depend sensitively on the control scheme  
adopted. 
This is 
quantified by implementing a shortcut to adiabaticity~\cite{RevModPhys.91.045001}, which we find can suppress non-adiabatic transitions even in the presence of strong environmental interactions, qualitatively modifying the WPD 
in ways that  
are often invisible at the level of  
the work mean or variance.
These results thus uncover genuine quantum features in the work statistics that are not captured by low-order 
moments but have a pronounced impact on the erasure fidelity, highlighting the 
need to access the WPD to fully characterise the thermodynamics of driven open quantum systems.

\section{Work Counting Statistics with Process Tensors}
\label{sec: work counting with PT}
\begin{figure*}
    \includegraphics[width=0.6\textwidth]{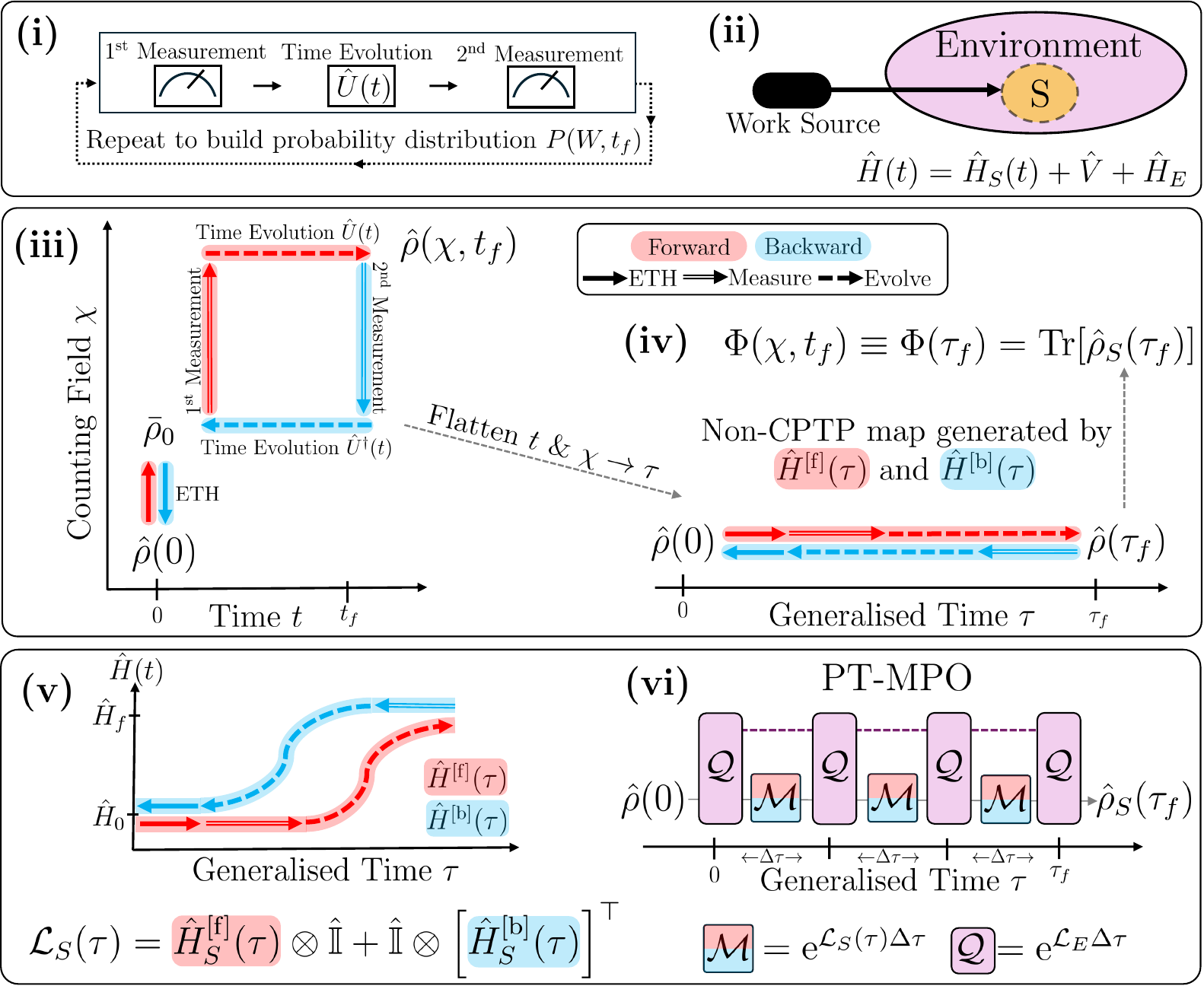}
    \caption{\textbf{Schematic of the process-tensor framework for quantum work statistics.}
(i) Two-point measurement protocol (TPMP) used to define the work probability distribution (WPD). 
(ii) A driven open quantum system $S$ exchanges energy and information with its surrounding environment $E$, governed by a time-dependent Hamiltonian $\hat{H}(t)$ under external driving from a work source. 
(iii) Calculation of the work characteristic operator (WCO), $\hat{\rho}(\chi,t_f)$, via propagation along both the physical time axis $t$ and the counting-field axis $\chi$. 
(iv) Mapping the $t$ and $\chi$ axes onto a single generalised-time axis $\tau$ allows the WCO (now $\hat{\rho}(\tau)$) to be evolved under a non-completely-positive, trace-preserving (non-CPTP) map generated by appropriately defined forward $\hat{H}^{[f]}(\tau)$ and backward $\hat{H}^{[b]}(\tau)$ generalised-time-dependent Hamiltonians. 
(v) The forward and backward Hamiltonians are constructed from the time-dependent Hamiltonian $\hat{H}(t)$; their system components form the generalised-time-dependent Liouvillian $\mathcal{L}_S(\tau)$. 
(vi) For an open quantum system, propagation along the $\tau$ axis is performed using process-tensor matrix-product-operator (PT-MPO) techniques, which yield numerically exact dynamics. 
All work-counting complexity is contained within the system-local propagators $\mathcal{M}$, while the $\mathcal{Q}$ tensors, connected through inner-bonds that carry non-Markovian correlations, encode the environmental influence functional.}
    \label{fig: TPMP, FWBW Hamiltonians}
\end{figure*}

Fig.~\ref{fig: TPMP, FWBW Hamiltonians} outlines our approach to overcoming the difficulties faced in calculating numerically exact open system work statistics 
by leveraging a combination of the process-tensor formalism,   
the TPMP,  
and a mapping to a generalised time axis. 
In the TPMP, projective measurements of a physical observable are assumed to be performed at times $0$ and $t_f$, and the difference between their outcomes defines the stochastic variable of interest (Fig.~\ref{fig: TPMP, FWBW Hamiltonians} (i)). 
For the work statistics of an externally driven open quantum system (Fig.~\ref{fig: TPMP, FWBW Hamiltonians} (ii)), these projective measurements must act on the  
joint system-environment state, since it is 
this which evolves unitarily according to the time-dependent Hamiltonian, 
\begin{equation}\label{totalH}
    \hat{H}(t) = \hat{H}_S(t) + \hat{V} + \hat{H}_E.
\end{equation}
Here, $\hat{H}_S(t)$ is the controlled system Hamiltonian, $\hat{V}$ is the system-environment interaction, and $\hat{H}_E$ describes the environment.

The TPMP  
then proceeds as follows:  
(1) prepare an initial composite state $\hat{\rho}_0$;  
(2) perform a projective energy measurement in the eigenbasis of $\hat{H}(0) \equiv \hat{H}_0$;  
(3) let the composite system evolve unitarily for a duration $t_f$;  
(4) perform a second projective energy measurement in the eigenbasis of $\hat{H}(t_f) \equiv \hat{H}_f$;  
(5) define the difference of the two outcomes as the stochastic work $W$;  
(6) repeat this procedure many times to construct the work probability distribution $P(W,t_f)$.

Although, in principle, this procedure yields the WPD for any quantum process, explicitly resolving the full energy spectrum of the combined system and environment is generally infeasible.  Instead, it is more convenient to evaluate the WPD via its Fourier transform, the \emph{work characteristic function}
\begin{align}
    \Phi(\chi, t_{f}) = \int_{-\infty}^{\infty} \dd W P(W, t_{f}) \mathrm{e}^{-\mathrm{i} \chi W},
\end{align}
where $\chi$ is a \emph{counting field}, here the conjugate variable to work transfer, and therefore time-like in nature. 
The WCF encodes the complete statistics of work: its inverse Fourier transform yields the WPD, while its derivatives give the moments directly, 
$\langle W^{n} \rangle (t_{f}) = (-\mathrm{i})^{n} {\dd_{\chi}^{n}} \Phi(\chi,t_{f}) |_{\chi=0}.$ 
It is also experimentally measurable~\cite{PhysRevLett.110.230601,DeChiara2018,oftelie2025measurementworkstatisticsopen}.

As shown in Ref.~\cite{talkner2007fluctuation}, the WCF for the TPMP can be written as
\begin{align}
\label{eq: WCF}
    \Phi(\chi, t_{f}) = \Tr[\mathrm{e}^{\mathrm{i}\chi \hat{H}_{f}} \hat{U}(t_{f})\mathrm{e}^{-\mathrm{i}\chi \hat{H}_{0}} \bar{\rho}_{0}\hat{U}^{\dagger}(t_{f})],
\end{align}
where $\hat{U}(t_{f}) = \mathcal{T}_{\leftarrow}\mathrm{e}^{-\ii\int_{0}^{t_{f}}\dd s \hat{H}(s)}$ is the time-ordered evolution operator, and $\bar{\rho}_{0}$ is the initial state of the composite system averaged over the first measurement, i.e.~dephased in the eigenbasis of $\hat{H}_{0}$. 
In our formulation, $\bar{\rho}_0$ is obtained by propagating the initial system-environment state $\hat{\rho}_{0}$ under the Hamiltonian $\hat{H}_{0}$ for an \emph{equilibration time} $t_{\mathrm{e}}$, such that $\bar{\rho}_{0}=\mathrm{e}^{-\mathrm{i}t_{\mathrm{e}} \hat{H}_{0}}\hat{\rho}_{0}\mathrm{e}^{\mathrm{i} t_{\mathrm{e}}\hat{H}_{0}}$. This induces dephasing in the energy eigenbasis, consistent with the eigenstate-thermalisation hypothesis~\cite{deutsch2018eigenstate}.

Using the cyclic property of the trace, we may write $\Phi(\chi, t_{f}) = \Tr[\hat{\rho}(\chi, t_{f})]$, where we define the \emph{work characteristic operator} (WCO)
\begin{align}
    \label{eq: WCO}
    \hat{\rho}(\chi, t_{f}) &= \hat{U}(t_{f})\mathrm{e}^{-\mathrm{i}\chi \hat{H}_{0}} \mathrm{e}^{-\mathrm{i}t_{\mathrm{e}} \hat{H}_{0}}\hat{\rho}_{0}\mathrm{e}^{\mathrm{i} t_{\mathrm{e}}\hat{H}_{0}}\hat{U}^{\dagger}(t_{f})\mathrm{e}^{\mathrm{i}\chi \hat{H}_{f}}.
\end{align}
This object is a pseudo density operator that evolves under a non-completely-positive, trace preserving (non-CPTP) map. 
Each exponential factor can be expressed as a time- or path-ordered integral, $\mathrm{e}^{\pm\mathrm{i} \hat{H}_{0}t_{\mathrm{e}}} = \mathrm{e}^{\pm\mathrm{i}\int_{c_{0}}\dd s \hat{H}_{0}}$ with $c_{0}\colon s\in [0, t_{\mathrm{e}}]$, $\hat{U}(t_{f}) = \mathcal{T}_{\leftarrow}\mathrm{e}^{-\mathrm{i}\int_{c_{1}}\dd {s} \hat{H}(s)}$ with $c_{1}\colon s\in [0, t_{f}]$, $\mathrm{e}^{-\mathrm{i}\chi \hat{H}_{0}} =\mathrm{e}^{-\mathrm{i}\int_{c_{3}}\dd s \hat{H}_{0}}$, and $\mathrm{e}^{\mathrm{i}\chi \hat{H}_{f}} =\mathrm{e}^{\mathrm{i}\int_{c_{3}}\dd s \hat{H}_{f}}$ with $c_{3}\colon s\in [0, \chi]$. 
We define operators with negative exponents as corresponding to \emph{forward} propagation, and those with positive exponents to \emph{backward} propagation. Note that this terminology should not be confused with that used in the context of fluctuation theorems~\cite{RevModPhys.81.1665,RevModPhys.83.771}. 
A schematic representing propagation along the time and counting field axes is shown in Fig.~\ref{fig: TPMP, FWBW Hamiltonians} (iii).

A key insight enabling our method is that, by concatenating these contours and defining forward and backward propagating Hamiltonians, the WCO can be evolved along a \textit{single generalised-time axis} $\tau$ 
\cite{funo2018path,Cavina23}. This mapping unifies the physical time and counting-field dynamics, transforming the multi-contour evolution into a one-dimensional propagation problem (see Fig.~\ref{fig: TPMP, FWBW Hamiltonians} (iv)). 
We shall see that this step allows the WCO to be expressed in terms of a process tensor~\cite{Pollock18}, making it possible to compute the full work statistics in a \textit{numerically exact and computationally efficient} manner. 
Mathematically, concatenating the contours gives
\begin{align}\label{eq:WCO}
    \hat{\rho}(\chi, t_{f})\equiv\hat{\rho}(\tau_{f}) &= \mathbb{T}_{\leftarrow}\mathrm{e}^{-\mathrm{i}\int\dd {\tau} \hat{H}^{\mathrm{[f]}}(\tau)}\hat{\rho}_{0}\mathbb{T}_{\rightarrow}\mathrm{e}^{\mathrm{i}\int\dd {\tau} \hat{H}^{\mathrm{[b]}}(\tau)} \notag \\
    & 
    = \hat{F}(\tau_{f})\hat{\rho}_{0}\hat{B}^{\dagger}(\tau_{f}),
\end{align}
where $\tau_f = t_e + \chi + t_f$. Propagation along the generalised-time axis is generated by the forward and backward unitary operators $\hat{F}(\tau)$ and $\hat{B}^\dagger(\tau)$, determined by the corresponding generalised-time-dependent Hamiltonians $\hat{H}^{[f]}(\tau)$ and $\hat{H}^{[b]}(\tau)$ (Fig.~\ref{fig: TPMP, FWBW Hamiltonians} (v)). 
The superoperator $\mathbb{T}_{\leftarrow}$ ($\mathbb{T}_{\rightarrow}$) orders operators in descending (ascending) order of $\tau$. 
Since $\hat{F}(\tau_{f})\hat{B}^{\dagger}(\tau_{f})\neq\mathbb{I}$, the propagation is non-CPTP. 
Taking the partial trace over the environment yields the WCF through $\Phi(\chi, t_{f}) \equiv \Phi(\tau_{f}) = \Tr_S[\hat{\rho}_{S}(\tau_{f})]$; 
thus, solving the reduced-system dynamics of $\hat{\rho}_S(\tau_f)$ provides numerically exact access to the complete work statistics of a driven open quantum system protocol.

Applying Eq.~(\ref{eq:WCO}) is intractable in its full form, as it requires diagonalising the full system-environment Hamiltonian. To overcome this, we exploit recent advances in matrix product operator (MPO) techniques for open-system dynamics~\cite{jorgensen2019exploiting,cygorek2022ACE,Cygorek24,Link2024infinite}. We discretise the generalised-time axis into steps of size $\Delta\tau$ and apply a Trotter decomposition to the forward and backward unitaries~\footnote{Note that since both the time and counting axes are mapped onto the $\tau$ axis, the physical time axis and the counting axis will have the same resolution.}. We then define 
$t_{\mathrm{e}} = s\Delta\tau$, 
$\chi = m\Delta\tau$, and 
$t_{f} = f\Delta\tau$, where $s$, $m$, and $f$ are integers whose sum $N=s+m+f$ gives the total number of propagation steps along $\tau$. 
Rewriting the WCO in terms of superoperators and employing this Trotter splitting yields 
\begin{align}
\label{eq: WCF PT}
    \hat{\rho}(\tau_{f}) \approx \prod_{j=1}^{N}\mathrm{e}^{\mathcal{L}_{E}\Delta\tau}\mathrm{e}^{\mathcal{L}_{S}(\tau_{j})\Delta\tau}(\hat{\rho}_{S}(0)\otimes\hat{\rho}_{E}(0)), 
\end{align}
with the system Liouvillian 
\begin{align}
\label{eq: counting Lio}
    \mathcal{L}_{S}(\tau_{j}) = \hat{H}^{[\mathrm{f}]}_{S}(\tau_{j})\otimes \hat{\mathbb{I}} + \hat{\mathbb{I}}\otimes [\hat{H}^{[\mathrm{b}]}_{S}(\tau_{j})]^{T},
\end{align}
where $\hat{H}^{[\mathrm{f}]}_{S}$ and $\hat{H}^{[\mathrm{b}]}_{S}$ are the system parts of the forward and backward generalised-time-dependent Hamiltonians, respectively.

Although the system Liouvillian $\mathcal{L}_S$ generates non-CPTP dynamics, Eq.~\eqref{eq: WCF PT} still corresponds to a discrete-time path integral~\cite{feynman1963theory} along the generalised-time axis. 
The central insight of this work is that the WCO can be obtained straightforwardly within any of the modern tensor-network representation techniques for real-time path integrals. These include time-evolving matrix product operators (TEMPO) \cite{strathearn_efficient_2018}, which is an MPO representation of QUAPI \cite{makri95theory, makri1995tensor}, or process tensor (PT-MPO) techniques~\cite{jorgensen2019exploiting,cygorek2022ACE,cygorek2024ACEcode,Cygorek24}, which represent the Feynman-Vernon influence functional in a compact MPO form (Fig.~\ref{fig: TPMP, FWBW Hamiltonians} (vi)). 
The PT-MPO formalism can be understood as compressing the propagator $e^{\mathcal{L}_E\Delta \tau}$ for the environment Liouvillian $\mathcal{L}_E$ via $\mathcal{Q}=\mathcal{C}e^{\mathcal{L}_E\Delta \tau}\mathcal{C}^{-1}$ onto the most relevant subspace of the full environment Liouville space~\cite{cygorek2025innerbonds}. This subspace is implicitly identified by optimality conditions for tensor network compression; the compression matrices $\mathcal{C}$ and their pseudo-inverses $\mathcal{C}^{-1}$ arise from truncated singular value decompositions (SVDs), where small singular values below a fixed compression threshold are dropped~\cite{orus2014practical}.
Since in our case the overall propagation time along the generalised-time axis is much longer than the memory time of the environment, the uniTEMPO algorithm~\cite{Link2024infinite}, which leverages time-translation invariance of the PT-MPO for spin-boson models, is particularly efficient. It is thus used throughout this article.

The resulting WCF is sensitive to both the equilibration and protocol durations, set by the integers $s$ and $f$, respectively, for a given timestep $\Delta\tau$. 
The resolution of the WPD extracted from the WCF is determined by the number of counting-field samples, which is chosen to be in the interval $m\in[0,\chi_{\mathrm{max}}/\Delta\tau]$. 

\section{Numerically Exact Work Statistics of Landauer Erasure}

Having established a general and numerically exact framework for calculating quantum work statistics in driven open systems, we now demonstrate its scope and flexibility by applying it to a paradigmatic information-to-energy process: Landauer erasure 
of a qubit~\cite{PhysRevLett.120.210601,gaudenzi_quantum_2018}. We show that 
resolving the full work probability distribution reveals distinctive features of the protocol while remaining fully tractable beyond the usual slow-driving and weak-coupling limits. 
The model considered is that of an externally driven two-level system (TLS) linearly coupled to a thermal environment of bosons, i.e.~the driven spin-boson model. The total Hamiltonian is of the form given in Eq.~(\ref{totalH}), 
with $\hat{H}_{E} = \sum_{k}\omega_{k}\hat{b}_{k}^{\dagger}\hat{b}_{k}$ defining the bosonic environment and $\hat{V} = \hat{\sigma}_{z}\otimes\sum_{k}(g_{k}^*\hat{b}_{k}^{\dagger}+g_{k}\hat{b}_{k})$ the system-environment interactions. Here, $\hat{b}_{k}$ annihilates an environmental boson of frequency $\omega_k$ that couples to the system with strength $g_k$. The initial composite state is taken to be separable, $\hat{\rho}_{0} = \hat{\rho}_{S}(0)\otimes\hat{\rho}_{E}$, with $\hat{\rho}_{S}(0)$ the maximally mixed state to be erased and $\hat{\rho}_{E}$ the Gibbs state of the environment at inverse temperature $\beta = 1/(k_{\mathrm{B}}T)$.  

\label{sec: qubit erasure results}
\begin{figure}
    \centering
    \includegraphics[width=0.999\linewidth]{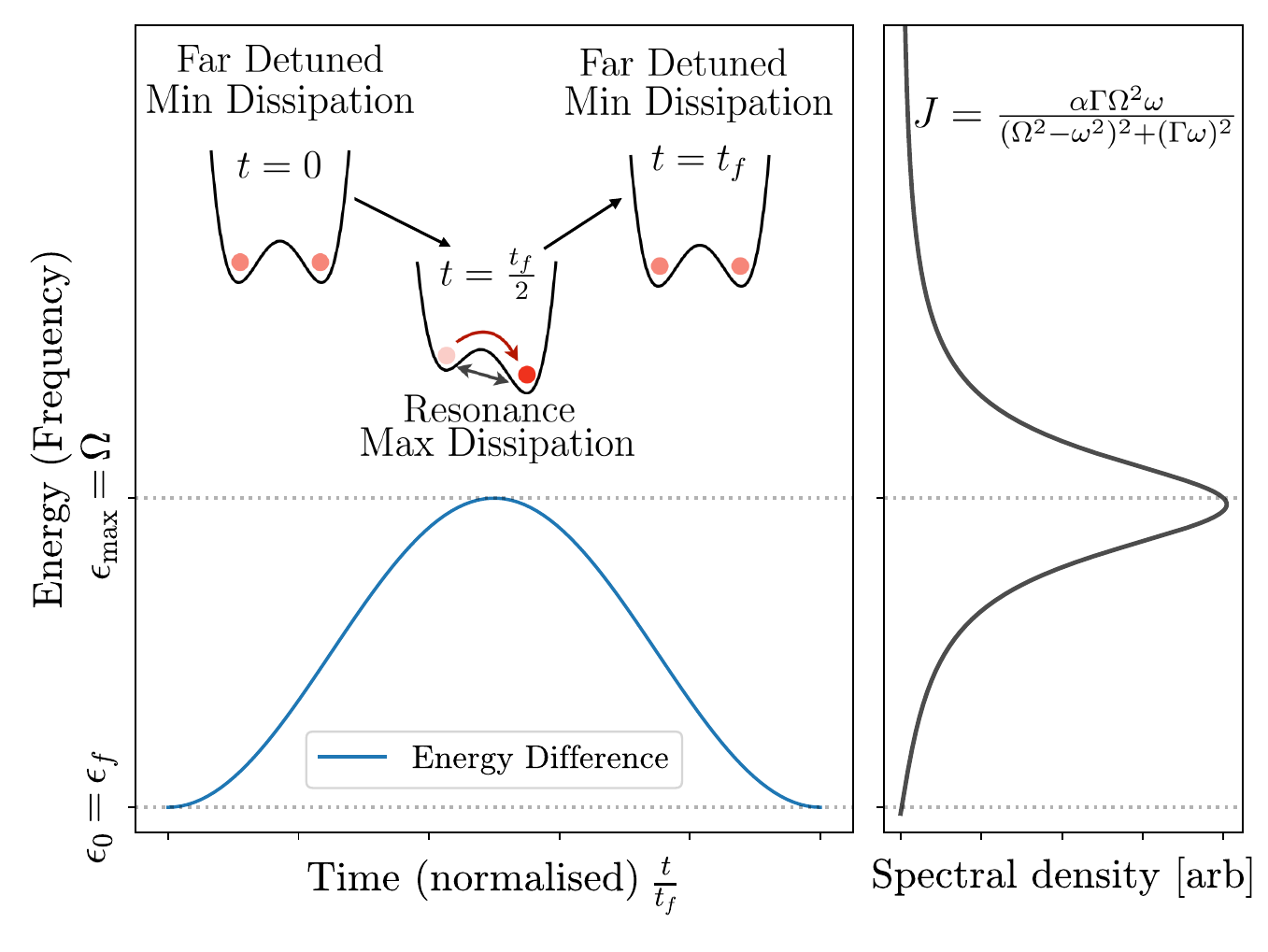}
    \caption{\textbf{Qubit erasure protocol.} The TLS energy splitting begins far from the peak of the spectral density ($\epsilon_0\approx 0 \ll \Omega$), is brought close to resonance with the peak halfway through the protocol ($\epsilon_{\mathrm{max}} \approx \Omega$), and is then returned to its initial far-detuned value ($\epsilon_{f} = \epsilon_{0} \ll \Omega$). 
    This cyclic design enables a thermodynamically consistent analysis of the work statistics. Throughout this work we use a maximum TLS energy splitting of $\epsilon_{\mathrm{max}} = 25\beta^{-1}$, an initial splitting of $\epsilon_{0} = 0.02\epsilon_{\mathrm{max}}$, and fixed environmental parameters $\Gamma=10\beta^{-1}$ and $\Omega=\epsilon_{\mathrm{max}}$. The coupling strength $\alpha$ and protocol duration $t_{f}$ are varied in subsequent figures.}
    \label{fig:energies}
\end{figure}

Erasure corresponds to driving the TLS from the maximally mixed state to (as close as possible) its ground state, thereby achieving the maximum possible reduction in system entropy. We implement this by driving the TLS over the interval $t \in [0,t_f]$ according to~\cite{miller2020quantum}
\begin{align}
\label{eq: erasure hamiltonian}
    \hat{H}_{S}(t) = \frac{\epsilon(t)}{2}\left[\cos\theta(t)\hat{\sigma}_z + \sin\theta(t)\hat{\sigma}_x\right],
\end{align}
with time-dependent energy splitting $\epsilon(t) = \epsilon_{0} + [\epsilon_{\mathrm{max}} - \epsilon_{0}]\sin^{2}(\pi t/t_{f})
$, and mixing angle $\theta(t) = \pi(t/t_{f} - 1)$~\footnote{We also add a shift term which ensures that the ground state energy is zero at all times.}. To satisfy the conditions for erasure we require $\epsilon_{0}\approx 0$ and $\epsilon_{\mathrm{max}} \gg \beta^{-1}$~\cite{miller2020quantum}.

\begin{figure*}[t]
    \centering
    \includegraphics[width=\textwidth]{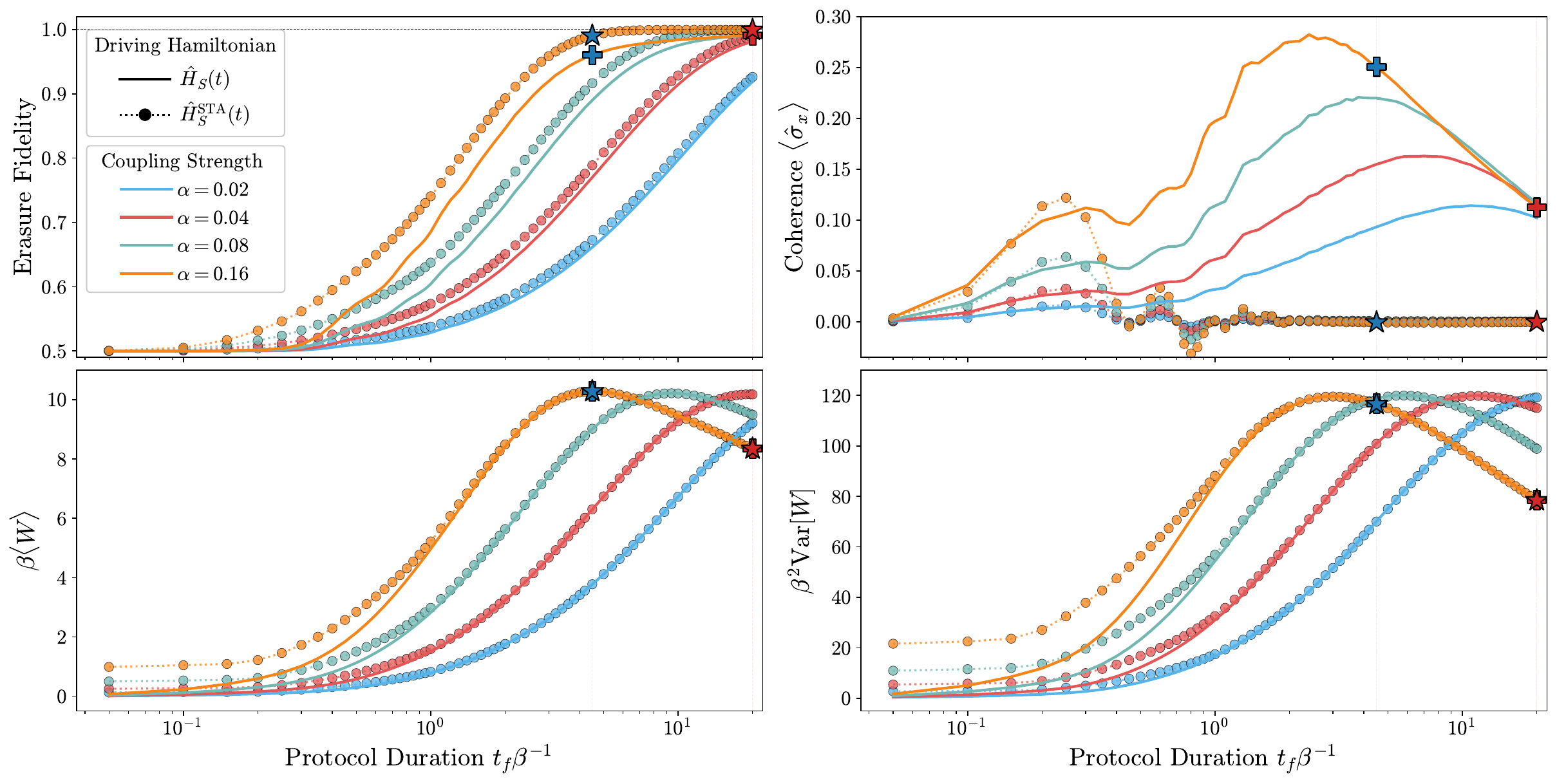}
    \caption{\textbf{Erasure fidelity, coherence, and low-order work statistics as functions of protocol duration.} 
    (Top left) Erasure fidelity, defined as the overlap between the final reduced system state and the ground state of $\hat{H}_S(t_f)$. 
    (Top right) TLS coherence in the energy eigenbasis of $\hat{H}_S(t_f)$, given by $\langle\hat{\sigma}_x\rangle$. (Bottom left) Mean work transfer. (Bottom right) Variance of work transfer. All quantities are evaluated at the end of the protocol and plotted against the protocol duration $t_f$ (logarithmic scale). Results are shown for different system-environment coupling strengths $\alpha$ (colours), for both the reference driving Hamiltonian in Eq.~\eqref{eq: erasure hamiltonian} (solid lines) and the STA-assisted protocol in Eq.~\eqref{eq: H STA} (circles, dotted lines). A generalised-time step of $\Delta\tau = 0.01$, SVD threshold $10^{-10}$, memory time $t_\text{mem}=5\beta$, and equilibration time $t_\mathrm{e} = 5\beta$ were used to ensure convergence. A tenth-order finite-difference method was employed to compute the work moments. 
    Selected points (stars and crosses) mark the protocols for which the full work probability distributions are analysed in Fig.~\ref{fig: wpd}.}
    \label{fig:dynamics-moments}
\end{figure*}

The formalism developed in Sec.~\ref{sec: work counting with PT} makes no adiabatic assumption, allowing us to explore non-adiabatic driving while retaining numerically exact access to the work statistics. Furthermore, because non-adiabatic 
transitions generally reduce the fidelity of erasure, adding a shortcut to adiabaticity (STA) might be expected to improve the overlap with the target ground state~\cite{RevModPhys.91.045001}. 
Defining STAs for open quantum systems is a difficult task, particularly beyond Lindblad master equations~\cite{Alipour2020shortcutsto,PhysRevLett.127.150401}. 
Hence, we study the efficacy of augmenting the erasure protocol with 
a closed system STA, which for the Hamiltonian in Eq.~\eqref{eq: erasure hamiltonian} results in
\begin{align}
    \label{eq: H STA}
    \hat{H}_{S}^{\mathrm{STA}}(t) = \hat{H}_{S}(t) + \frac{\pi}{2t_{f}}\hat{\sigma}_{y},
\end{align}
as shown in Appendix~\ref{sec: STA}. For consistency with the two-point measurement protocol, the STA is switched on (off) immediately after (before) the initial (final) projective energy measurement. Later, we compare erasure fidelities, low-order work moments, and full work distributions for protocols with and without the STA, which we find to be effective despite being defined for a closed system.

We consider an environment modelled by an under-damped Drude-Lorentz spectral density, $J(\omega) ={\alpha\Gamma\Omega^{2}\omega/[(\Omega^{2}-\omega^{2})^{2} + (\Gamma\omega)^{2}}]$, representing a reservoir of width~$\Gamma$ centred at~$\Omega$,  with system coupling strength~$\alpha$. 
The time-dependent protocol of Eq.~\eqref{eq: erasure hamiltonian} is illustrated in Fig.~\ref{fig:energies}. 
The TLS energy splitting starts far below the spectral density peak, $\epsilon_{0} \ll \Omega$. In this regime, the TLS and bath interact only weakly, leading to minimal heat dissipation. As a result, erasure does not occur unless the system is actively driven. The energy splitting is then increased until the midpoint of the protocol, $t=\frac{t_{f}}{2}$, where the TLS splitting is close to resonant with the spectral density peak, $\epsilon_{\text{max}}=\Omega$, and the coupling is consequently strongest. 
The protocol is symmetric and cyclic (the latter ensuring thermodynamic consistency), so in the second half the energy splitting is brought back away from resonance to its initial value, $\epsilon_{f} = \epsilon_{0} \ll \Omega$. This design allows the qubit to operate in a regime of weak effective coupling, while enabling erasure by tuning the splitting close to resonance with the bath when required.

\subsection{Erasure fidelity and work moments}
\label{sec: fidelity and moments}

To characterise the performance of the protocol under both non-adiabatic driving and finite system-environment coupling, we begin by examining the erasure fidelity, TLS coherence, and the lowest-order work moments. Fig.~\ref{fig:dynamics-moments} 
displays these quantities evaluated at the end of the protocol as a function of the protocol duration (horizontal axis, logarithmic scale). The erasure fidelity is defined as the overlap of the reduced system state with the ground state of $H_{S}(t_{f})$, and the TLS coherence is measured via $\langle\hat{\sigma}_x\rangle$ since $\hat{H}_S(t_f)=(\epsilon_0/2)\hat{\sigma}_z$. Results are shown for a range of coupling strengths, under both the reference driving Hamiltonian in Eq.~\eqref{eq: erasure hamiltonian} (solid lines) and the STA-assisted Hamiltonian in Eq.~\eqref{eq: H STA} (circles, dotted lines).

Erasure fidelity improves in three clear regimes: for longer protocol durations (i.e.,~slower driving), for stronger system-environment coupling, and when the STA is included. 
As well as suppressing  non-adiabatic transitions, longer protocols keep the TLS near resonance with the spectral density peak for an extended period, increasing the probability of dissipating heat into the bath and thereby completing the erasure. Stronger coupling enhances this effect, particularly for short to intermediate protocol times, where the qubit would otherwise have insufficient time to dissipate the required heat. In essence, the degree of erasure depends on the interplay between the instantaneous effective coupling strength and the time spent in stronger effective coupling regimes: strong coupling can compensate for a short protocol, whereas longer protocols are required when the coupling is weaker. The improvement due to the STA is most pronounced at stronger couplings and intermediate (non-adiabatic) protocol durations 
and can be attributed to adiabatic undressing~\cite{adiabatic_undressing}: even when strong coupling to the bath leads to complete population of the instantaneous ground state approaching the peak of the drive ($t\approx t_f/2$), the drive has to be turned off slowly enough in the interval $t\in[t_f/2, t_f]$ for the system to follow the instantaneous ground state all the way to $t=t_f$. The STA is beneficial for shorter protocol times $t_f$ as the undressing dynamics remains adiabatic even for faster drives.

Interestingly, while the STA leads to noticeable differences in erasure fidelity for moderate protocol durations, the low-order work moments do not exhibit an equally clear distinction. If one were to examine only the mean and variance of the work, it would be easy to overlook the impact of the STA on the underlying work statistics, despite its marked effect on erasure fidelity and coherence. To probe this discrepancy, in the following section we compute the full work probability distribution for two representative protocols indicated by the coloured stars and crosses in Fig.~\ref{fig:dynamics-moments}. 
These particular protocols were chosen because they exhibit a pronounced difference in their erasure fidelities and final coherences, while displaying very similar mean and variance of the work.

Finally, we note two qualitative features visible in Fig.~\ref{fig:dynamics-moments}. First, both the mean and variance of work display a non-monotonic dependence on $t_f$, rising to a maximum at intermediate durations before decreasing again.  
Second, for very short protocol durations, the mean and variance remain finite for STA-assisted driving even though erasure becomes ineffective. Both of these features are explained by considering the full WPD, which we do in the next section.

\subsection{Full statistics: work probability distribution}
\begin{figure}
    \centering
    \includegraphics[width=\linewidth]{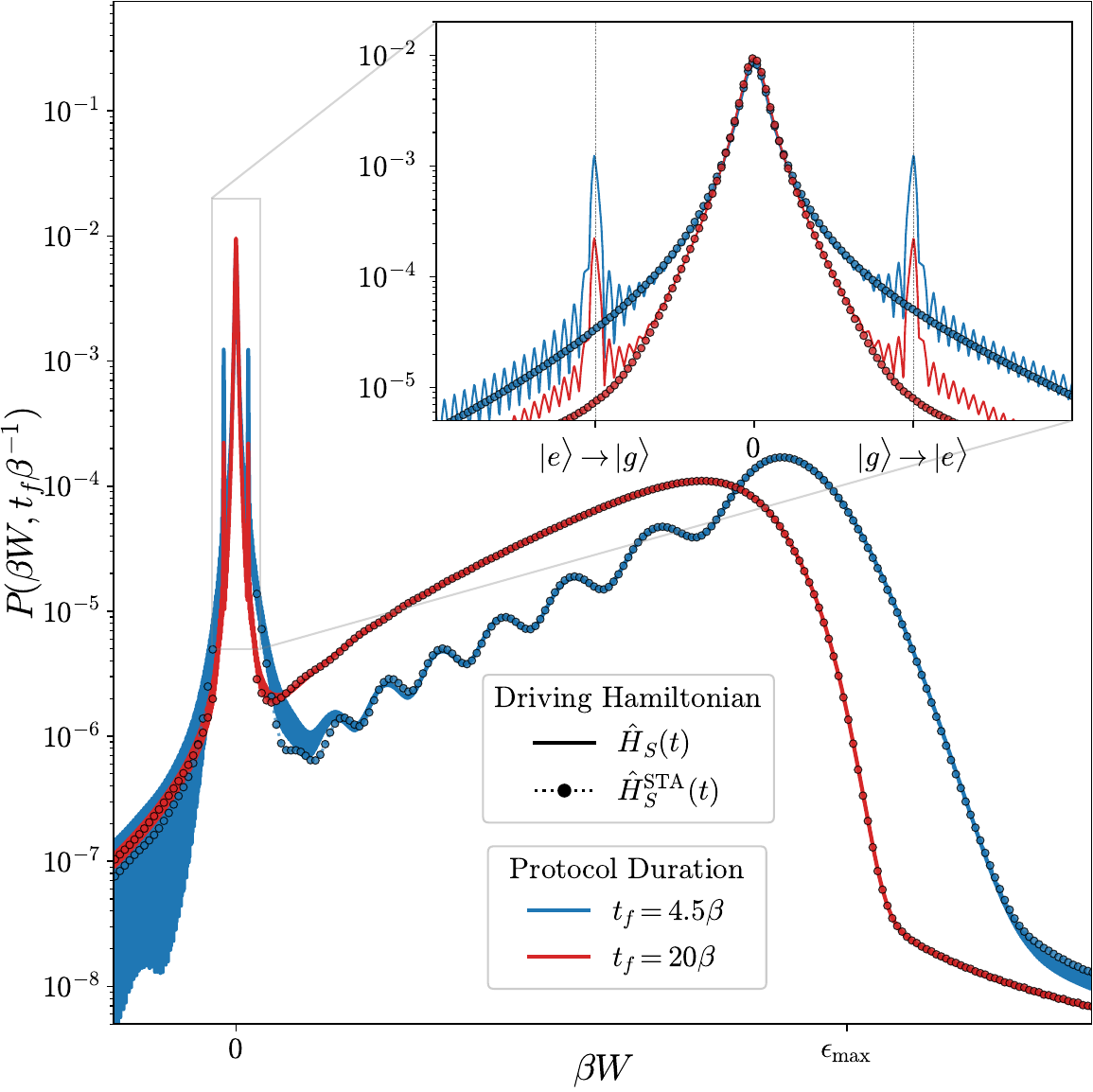}
    \caption{\textbf{Full work probability distributions for representative Landauer erasure protocols.} 
    Work probability distributions for coupling strength $\alpha = 0.16$ and protocol durations $t_f = 4.5\beta$ (blue) and $t_f = 20\beta$ (red).
    Results are shown for both the driving Hamiltonian $\hat{H}_S(t)$ (solid curves) and the STA-assisted Hamiltonian $\hat{H}^{\mathrm{STA}}_S(t)$ (circles). The main panel highlights the broad positive-work distribution accompanied by heat dissipation during the erasure process. The inset displays the region around $W=0$, where closed-system transitions manifest as distinct peaks and we see a clear suppression of non-adiabatic work values when using the STA. Notably, the STA has little effect on the rest of the WPD, for either protocol duration. The WPDs shown here correspond to the highlighted points within Fig.~\ref{fig:dynamics-moments} (blue and red stars and crosses) and were obtained by Fourier transforming converged WCFs, calculated using a generalised-time step $\Delta\tau = 0.01$, SVD threshold $10^{-10}$, and equilibration time $t_{\mathrm{e}} = 5\beta$. The largest counting field value used was $\chi_{\mathrm{max}}=200\beta$, after which the real and imaginary parts of the WCF had decayed to less than $10^{-3}$. The distribution is calculated with a fast Fourier transform (FFT), binned with $\Delta W = 0.002\beta^{-1}$. A slow decaying envelope of $\exp[-0.005\chi]$ was added to the WCF before performing the FFT, smoothing out some of the fast oscillations present in the WPD, without changing its shape qualitatively.}
    \label{fig: wpd}
\end{figure}

To clarify the origin of the differences observed in Sec.~\ref{sec: fidelity and moments} between protocols that exhibit similar low-order work moments yet markedly different erasure fidelities and coherences, we now examine the complete WPD.
Fig.~\ref{fig: wpd} presents the WPD for the two representative erasure protocols indicated in Fig.~\ref{fig:dynamics-moments}, each with coupling strength $\alpha=0.16$, and with different protocol durations of $t_{f}=4.5\beta$ (blue) and $t_{f}=20\beta$ (red)~\footnote{For these WPDs, the WCFs were converged using a generalised-time step of $\Delta\tau = 0.01$, an SVD precision of $10^{-10}$, and an equilibration time of $t_{\mathrm{e}} = 5\beta$. After obtaining the WCF, we computed the full probability distribution using a fast Fourier transform with a bin width $\Delta W = 1/500$, together with an exponential damping factor $\exp[-0.005\chi]$ to ensure that the WCF decays to zero. 
This damping affects the results only by slightly broadening the peaks that appear in the work distribution.}. For both durations, we compare the results obtained by driving the qubit according to the Hamiltonian $\hat{H}_S(t)$ (solid lines) and the STA Hamiltonian $\hat{H}^{\mathrm{STA}}_S(t)$ (circles, dotted lines). The STA data are omitted near $W=0$ in the main panel for clarity, but are shown explicitly in the inset.

We first focus on the region close to $W=0$, shown in the inset. When the TLS is driven by $\hat{H}_{S}(t)$, the WPD exhibits three well-defined peaks arising from closed-system transitions that can occur during the TPMP, i.e.,~projections onto the system ground and excited states at the start (when the system state is maximally mixed) and the end of the erasure process without the exchange of energy with the environment. The two sharp peaks at symmetrically placed positive and negative work values correspond to the non-adiabatic transitions $\ket{g}\to\ket{e}$ and $\ket{e}\to\ket{g}$, respectively. The central peak at $W=0$ arises from projections onto the same energy eigenstate at both measurement times ($\ket{g}\to\ket{g}$ or $\ket{e}\to\ket{e}$). When the STA is applied, the side peaks associated with the non-adiabatic transitions are strongly suppressed for both protocol durations, while the central $W=0$ contribution remains essentially unchanged. 
This directly explains the differences in erasure fidelity observed in Fig.~\ref{fig:dynamics-moments}. Even for Landauer erasure in an open quantum system, the STA removes unwanted non-adiabatic pathways, thereby improving erasure fidelity, without materially affecting the remaining work distribution or its low-order moments.

Beyond the vicinity of $W=0$, the WPD is dominated by a broad feature at positive work, corresponding to the energetic cost of erasure due to heat dissipation into the bath. This feature provides further insight into the peaks seen in the work mean and variance in Fig.~\ref{fig:dynamics-moments}. The slower protocol ($t_f = 20\beta$) produces a distribution shifted to smaller work values, since heat is more likely to be dissipated by the TLS before its energy splitting reaches the spectral density peak and heat absorption also becomes more significant. Consequently, the mean and variance of work are lower than for the faster protocol. In contrast, for the more strongly non-adiabatic protocol ($t_f = 4.5\beta$) there is an increased chance of dissipation (and therefore work transfer) when the TLS splitting is closer to $\epsilon_{\max}$, shifting the work distribution to higher values and thereby increasing both the mean and variance. Reducing the protocol duration further the WPD becomes increasingly dominated by features arising from closed-system transitions near to $W=0$, leading to a reduction in entropy exchange and hence erasure fidelity. When driving with $H_{S}(t)$, the non-adiabatic side peaks eventually dominate for short enough protocol durations, whereas with the STA present these peaks are suppressed and the central $W=0$ peak remains along with a broad, though smaller in amplitude, dissipative feature. This leads to the non-vanishing mean and variance at small $t_f$ in the STA case.

Finally, we comment on the oscillations visible in the WPD for the shorter protocol. As shown in Appendix~\ref{sec: emission spectrum}, similar oscillations appear in the time-integrated emission spectrum of the TLS in this regime~\cite{moelbjerg2012,boos2024}, 
suggesting that they originate from interference of energy exchanges between dynamically dressed states at different points in time. 

\section{Conclusions}
\label{sec: conclusions}
We have developed a numerically-exact framework for calculating the full counting statistics of work transfer in driven open quantum systems. By introducing a generalised-time axis and constructing the corresponding forward and backward Hamiltonians, we obtain the WCF by propagating the WCO (a pseudo density operator that evolves under a non-CPTP map) along this axis. From the WCF, the work moments follow directly from derivatives, and the full probability distribution can be recovered via an inverse Fourier transform. The framework is numerically exact; given sufficient computational resources, the work statistics can be computed to arbitrary accuracy. No assumptions about the system-bath coupling strength nor the driving are required, enabling the study of strongly coupled, non-Markovian, and non-adiabatic regimes. 
The only requirement for Eq.~\eqref{eq: WCF PT} is that the environment Hamiltonian is time independent, and that calculating the PT-MPO to the desired accuracy is numerically feasible. 
We have illustrated the power of this approach using a physically motivated Landauer erasure protocol. Our results reveal rich structure in the full work distribution at strong environmental coupling  
and show how including an STA in the system Hamiltonian leads to modifications 
by suppressing peaks associated with non-adiabatic qubit transitions. This increases erasure fidelity, without impacting low-order work moments.

The general framework introduced here can be readily applied to calculate work statistics of other control protocols and system-environment interactions. For a fixed system-environment coupling, multiple driving strategies can be tested simply by redefining the forward and backward generalised-time-dependent Hamiltonians, enabling the same process tensor to be reused for efficient optimisation of work costs and other performance metrics. More broadly, the process-tensor formulation we have developed can be applied beyond the specific cases considered here, to bosonic, fermionic, or spin environments with arbitrary spectral densities and general time-dependent system Hamiltonians. While we have used the uniTEMPO algorithm to construct the work-counting process tensor, this is not a limitation; alternative tensor-network techniques, such as the ACE algorithm~\cite{cygorek2022ACE}, can be incorporated straightforwardly to extend the classes of systems that may be explored. Owing to its versatility, we expect the present framework to be valuable both for quantum control, where it can aid in identifying protocols that minimise work cost while maintaining high fidelity, and for quantum thermodynamics, where it enables access to strongly coupled and non-Markovian regimes that remain challenging for existing techniques.

\section{Acknowledgements}
MS acknowledges support from the EPSRC (EP/SO23607/1 and EP/W524347/1). M.C. acknowledges funding by the Return Programme of the State of North Rhine-Westphalia. We thank Valentin Link, Harry Miller, Mark Mitchison, and Paul Skrzypczyk for useful discussions.

\begin{appendix}

\begin{figure*}[t]
    \centering
    \includegraphics[width=\textwidth]{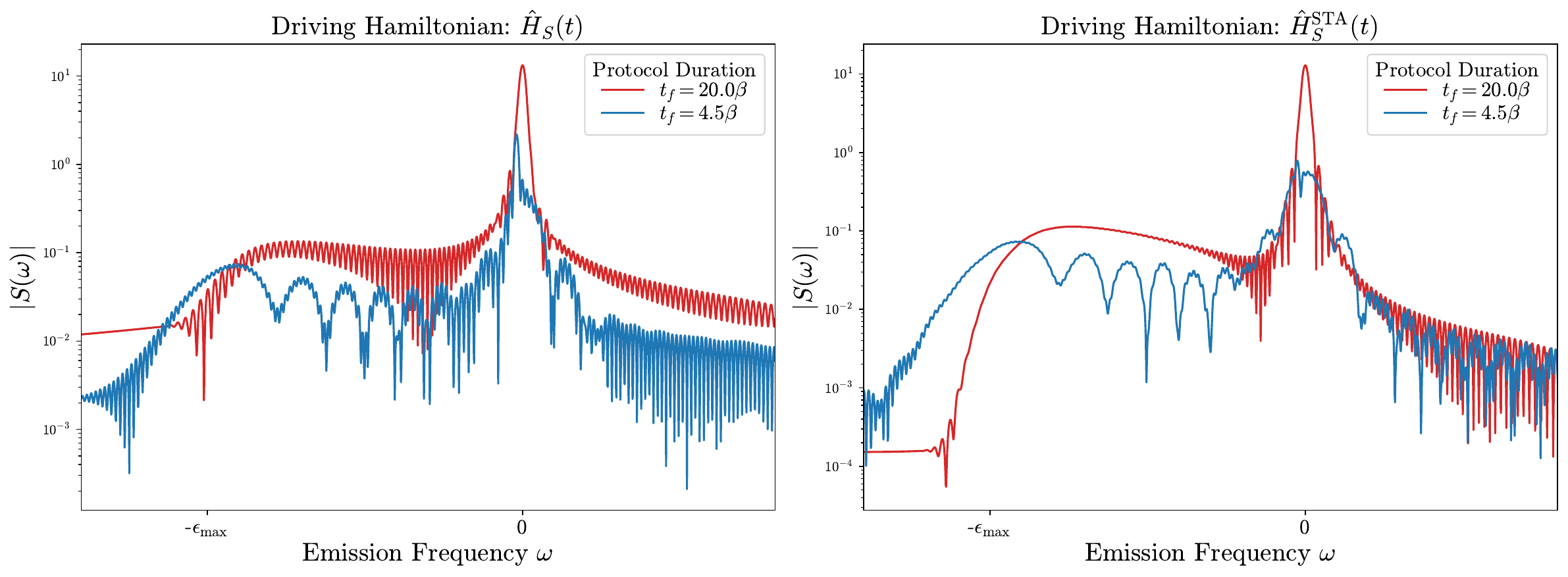}
    \caption{Time integrated emission spectra of the TLS during the Landauer erasure protocols presented in Fig.~\ref{fig: wpd} (left -- without STA, right -- with STA). We see similar oscillations in the emission spectra to those in the WPD for the faster protocol (blue), where the emission frequencies are, as expected, of the opposite sign to the work values in Fig.~\ref{fig: wpd}.}
    \label{fig:emission_spectrum}
\end{figure*}

\section{Shortcut to adiabaticity}
\label{sec: STA}
Here we derive the shortcut-to-adiabaticity (STA) term for the system Hamiltonian introduced in Eq.~\eqref{eq: erasure hamiltonian},
\begin{equation}
    \hat{H}_{S}(t) = \frac{\epsilon(t)}{2}\left[\cos\theta(t)\hat{\sigma}_z + \sin\theta(t)\hat{\sigma}_x\right].
\end{equation}
Consider a unitary transformation of the system state of the form
\begin{align}
    \hat{\rho}_{S}'(t) = \hat{V}^{\dagger}(t)\hat{\rho}_{S}(t)\hat{V}(t),
\end{align}
generated by
\begin{align}
    \hat{V}(t) = \mathrm{e}^{-\frac{\mathrm{i}\phi(t)}{2} \hat{\sigma}_{y}}.
\end{align}
The transformed system state obeys 
\begin{align}
    \dt\hat{\rho}_{S}'(t) = -\mathrm{i}\comm{\hat{H}_{S}'(t)}{\hat{\rho}_{S}'(t)},
\end{align}
with the transformed Hamiltonian
\begin{align}
    \hat{H}_{S}'(t) = \hat{V}^{\dagger}(t)\hat{H}_{S}(t)\hat{V}(t) - \frac{\dot{\phi}(t)}{2}\hat{\sigma}_{y},
\end{align}
where
\begin{align}
    \hat{V}^{\dagger}(t)\hat{H}_{S}(t)&\hat{V}(t)= \notag\\ &\frac{\epsilon(t)}{2}\bigg[
    \left\{\cos\theta(t)\cos\phi(t) + \sin\theta(t)\sin\phi(t)\right\}\hat{\sigma}_{z}\notag\\
    &+ \left\{\sin\theta(t)\cos\phi(t) - \cos\theta(t)\sin\phi(t)\right\}\hat{\sigma}_{x}\bigg].
\end{align}
To suppress non-adiabatic transitions, the coefficient of $\hat{\sigma}_{x}$ must vanish, which requires $\phi(t) = \theta(t)$. Substituting this choice gives the STA Hamiltonian
\begin{align}
    \hat{H}_{S}^{\mathrm{STA}}(t) = \hat{H}_{S}(t) + \frac{\dot{\theta}(t)}{2}\hat{\sigma}_{y} = \hat{H}_{S}(t) + \frac{\pi t_{f}}{2}\hat{\sigma}_{y},
\end{align}
such that 
\begin{align}
    \hat{H}_{S}'(t) =
    \frac{\epsilon(t)}{2}\hat{\sigma}_{z}.
\end{align}
When evaluating work statistics using the TPMP, the projective measurements must be performed in the eigenbasis of the original Hamiltonian $\hat{H}_{S}(t)$, rather than that including the counterdiabatic driving, $\hat{H}_{S}^{\mathrm{STA}}(t)$. Since the counterdiabatic term is constant in time in this example, it cannot be forced to vanish at both $t=0$ and $t=t_f$. We therefore assume that it is switched on quickly just after the first measurement and switched off quickly just before the second measurement. We have also tested smoothed switch functions, which lead to no qualitative changes to the results presented.

\section{Origin of oscillations in the work distribution}
\label{sec: emission spectrum}

In Fig.~\ref{fig: wpd}, we observed oscillatory modulations on top of the broad positive work distribution for short protocol durations. 
These oscillations are not directly associated with a single system parameter and their origin is therefore not immediately obvious. Qualitatively similar behaviour has been theoretically predicted~\cite{moelbjerg2012} and experimentally observed~\cite{boos2024} in the emission spectra of semiconductor quantum dots under pulsed excitation, where the effect was attributed to interference between emission peaks associated with the dot instantaneous energy eigenstates at different points in time. 

To investigate whether a comparable mechanism is responsible for the oscillations in the WPD, we compute an analogous quantity, the time-integrated emission spectrum of the two-level system
\begin{align}
    S(\omega) =  \int_{-\infty}^{\infty}\mathrm{d}s\mathrm{e}^{\mathrm{i}\omega s}\int_{0}^{t_{f}}\mathrm{d}t\langle \hat{\sigma}^{+}(t+s)\hat{\sigma}(t)\rangle.
\end{align}
The resulting spectra are shown in Fig.~\ref{fig:emission_spectrum}. Irrespective of the protocol time, we find fast oscillations in the emission spectra, which are however generally damped when the STA protocol is used. 
More importantly, we also find slower oscillations for the shorter protocol time, both with and without the STA, a feature that closely mirrors the behaviour of the WPDs in Fig.~\ref{fig: wpd}. 
This similarity suggests that the modulations observed in the WPDs also arise from interference effects associated with dynamically dressed states sampled at different points throughout the protocol.

\end{appendix}

\bibliographystyle{apsrev4-2}
\bibliography{bib.bib}

\end{document}